# Archean Paleo-climate: The first snowball?

Hector Javier Durand-Manterola

Departamento de Ciencias Espaciales, Instituto de Geofísica
Universidad Nacional Autónoma de México

E-mail: hdurand@igeofisica.unam.mx

Abstract

*Context:* The model accepted is one where during the Archean Eon the Earths climate was clement despite the weaker Sun. The observational evidence that supports this concept is: the emergence of life, the existence of evaporitic sediments and the presence of terrigenous sediments, all of which require liquid water and clement conditions. A theoretical argument used to support this idea is the so called ice-albedo feedback, which states that if the Earth was frozen, it would still be frozen.

*Objective:* The aim of this document is to present an alternative scenario in which a frozen world, "snowball" style, with liquid water at the bottom of the sea, also allows for the emergence of life and evaporitic and terrigenous sedimentation.

*Method:* Archean climatic evidence, available at present, is discussed and can be reinterpreted to support the idea that, in Archean times, the surface of the Earth was frozen. Also, a mathematical model is being developed to demonstrate that the ice-albedo feedback is not an inevitable consequence of a frozen Archean Eon.

*Results:* Reinterpretation of the evidence shows that life could appear within the oceanic depths and not necessarily on the surface. The evaporitic sediments could have formed by saline saturation of the water enclosed in the limited cavities of liquid water located at the bottom of the ocean. Also, the terrigenous sediments could have been formed by catastrophic currents of liquid water due to the fusion of the ice from the sub glacial volcanoes. From the mathematical model it is deduced that the defrosting moment of the Earth is towards the end of the Proterozoic, moment in which the evidence shows the "snowball" Earth ends.

Key words: Archean, paleo-climatology, paleo-environment

1. Introduction

Stellar evolution models establish that the Sun, during the Archean Eon, had less luminosity than in present day (Endal, 1980; Gough, 1981; Walter, 1982). The implications which arise from this are that, with this diminished luminosity and if the Earth had today's actual conditions, the planet should have been frozen (Sagan and Mullen, 1972; Kasting and Grinspoon, 1991). However, empirical evidence exists which seems to indicate that the Earth was free of ice during the Archean Eon. This contradiction is known as the Faint Young Sun Paradox (Crowley and North, 1991, pp. 238; Kasting and Grinspoon, 1991).

The empirical evidence which is provided in favor of an ice free Earth during the Archean Eon is as follows:

a) The existence of stromatolites from 3.6 Ga, ones which are formed by photoautotroph organisms which need liquid water and sunlight to live (Crowley and North, 1991, Walter, 1983), implying, therefore, that the surface of the ancient ocean should have been in a liquid state.

b) The existence of evaporitic sediments needed for the formation of liquid water with a free surface (Goodwin, 1991).

c) The production of terrigenous sediments during the early archaic, in other words, continental sediments produced by the erosion of running water, including shale, argillite, mudstone, quartz sandstone, and conglomerate (Goodwin, 1991).

As well as this observed evidence there is a theoretical argument supporting the idea of a mild climate during the Archean Eon, the so called ice-albedo feedback. This result was obtained using zero dimensional climatic models and demonstrates that if in the past the planet was covered with ice then in the present it would remain frozen, although the solar radiation has increased (Crowley and North, 1991, pp. 16). This is due to the fact that the ice has a very high albedo and hence very little solar heat is absorbed, whilst the ice is on the surface of the planet, maintaining the median temperature of the planet beneath freezing point. To solve the Faint Young Sun Paradox it has been proposed that the amount of greenhouse gases, like $CO_2$ and $NH_3$, were greater in the Archean Eon atmosphere than at present and this would produce a much greater greenhouse effect than the actual one and it maintained the temperature above freezing point (Sagan and Mullen 1972; Hart, 1978).

One problem which arises with this solution to the paradox is that it is not the only solution and that the little evidence in its favor can be reinterpreted in favor of a completely different model.

In this work I will show that an alternative scenario, with a planet covered with ice, it is possible to support it with the same empirical evidence simply by reinterpreting this evidence. It is showed that all climatic evidence from the Archean Eon, available at present, can support the idea that during the Archean times the surface of the Earth was frozen. The "snowball" Earth of the late Proterozoic (Hoffman et al., 1998) is not anything more than the end of that long period of freezing. The freezing of the Earth and its defrosting at the end of the Precambrian Era also explains the so called "explosion" during the Cambrian Period, in which life started to develop with a much greater variety than during previous eons. Arguments that demonstrate that on the frozen Earth there would not be free $CO_2$ in the atmosphere were also presented. Also an analytic model is presented to demonstrate that the ice-albedo feedback is not an inevitable consequence of a frozen Archaic Eon.

## 2. The frozen scenario

The basic idea of the frozen scenario is that the global ocean was frozen on the surface but at the bottom of the sea there was liquid water. During this time, like today, within the oceanic dorsals the fresh basaltic magma was exposed, and was cooled down by the hydrothermal circulation through the fractures. In these zones the heat flow, from inside Earth, kept liquid water near the seabed. Evidence that supports this idea is the existence of the Vostok Lake in Antarctica. This lake is a mass of fresh water with a length of 300 km and a depth of 400 m. It is located 3,700 m beneath the Antarctic ice layer. This lake was formed in a rift valley in which the heat flow maintains the water liquid (Carsey and Horvath, 1999). We can think of the Antarctic Ice and Vostok Lake as a scale model of the frozen Earth during the Archean and Proterozoic Eons.

## 3. Reinterpretation of the evidence

*3.1. Life and stromatolites.* The water which rises from the hydrothermal vents in the oceanic dorsals is rich in dissolved minerals (Sherwood et al.,

1983, pp. 86) and there is evidence that these types of hydrothermal vents already existed during the Archean (Kelley et al., 2005). With hot water and dissolved minerals, these zones have favorable conditions for the development of life. At present these zones are unusually rich in anaerobic bacteria, with a biomass 500 to 1000 times greater than in the surrounding areas (Rice and Lambshead, 1994 pp. 471). It has been suggested that life could have appeared in these zones and even that it was easier than on the surface (Maher and Stevenson, 1988; Pantoja-Alor and Gomez-Caballero, 2004; Kelley et al., 2005; Bradley, 2009). If life appeared in these deep zones the surface of the sea could have been frozen.

The best evidence of life in the Archean is the stromatolites. The stromatolites are produced by autotrophic organisms but it is not known if the archaic stromatolites were photoautotrophs or chemoautotrophs (Walter, 1983, pp. 192). If the organisms which formed stromatolites during the Archean were chemoautotrophs they could have grown at a great depth in the geothermal zones without light, or they could have been photoautotrophs, doing photosynthesis using infrared radiation, as they do today some bacteria in hydrothermal vents (White et al., 2002). Although all the present day stromatolites live in shallow waters, certain evidence exists that there existed deep water stromatolites during the Jurassic period (Gomez-Pérez, 2001). Hence the existence of stromatolites cannot be used as evidence stating that the surface of the global ocean was liquid.

*3.2 Evaporitic Sediments.* With respect to the evaporitic sediments found on Archean rocks, these types of sediments are formed when the substance dissolved in water reaches its saturation point and then it is deposited. However, it is not necessary that this saturation occurs due to evaporation in inland areas or in shallow waters. The saturation point can also be reached by the continuous dissolving of the substance in a limited amount of water. With the frozen Earth model this is the case because the liquid water zones are closed and limited (think in the Vostok Lake) and the submarine hydrothermal vents where continuously contributing with dissolved substances. Hence the existence of evaporites is also not evidence of a defrosted marine surface or of continents with lakes of liquid water.

*3.3 Terrigenous Sediments.* Some terrigenous sediments (eg. Greywacke) are found in Archean rocks (Budyko et al., 1985). In a frozen environment the terrigenous sediments should not have formed but could have been produced by the eruption of the volcanoes underneath the icy layer. Melting large quantities of ice (several kilometers thick) the torrent liberated could erode the terrain as it has done by the volcanoes located under the ice in Iceland and the Antarctic (Corr y Vaughan, 2008). Hessler y Lowe (2006) find evidence of an aggressive weathering environment during the Archean Eon. This should be expected with huge quantities of water flowing violently by volcanic melting.

Another possibility is that these "terrigenous materials", due to the closeness of the Moon, were generated by the erosion of the marine floor due to the submarine currents created by the gigantic tides (~ 350 m hight), during this period. A proto-continent submerged at a lesser depth than 350 m will receive very strong erosion in the moment in which the tide wave rides over it.

Another remote possibility is that normal submarine currents which present a greater flow than the flow from the bigger rivers (Whitehead, 1989; Worthington, 1969; Smith, 1975, Whitehead and Worthington, 1982) would have eroded the submerged continental type rocks producing pseudo terrigenous sediments in the same way that the rivers erode the earth.

## 4. Lack of $CO_2$ in the atmosphere

It is possible to doubt about the existence of large quantities of $CO_2$ in the atmosphere during the Archean period because of several reasons:

*4.1 Water as a solvent.* Water is an efficient solvent of $CO_2$ and, in the absence of well developed continents, the volcanism was purely submarine and it did not emit gases into the atmosphere because they dissolved directly into the water (Sherwood et al., 1983, pp. 86). When the $CO_2$ dissolves in water it forms carbonic acid and when it reacts with the magnesium and calcium ions dissolved in the ocean it forms carbonates which then get deposited on the seabed, eliminating the dissolved $CO_2$. Stromatolites builders' organisms would have also contributed to this process. At present corals and cyanobacteria are the main removers of $CO_2$ in sea water.

*4.2. Clathrates.* Another reason to think that the atmosphere during the Archaic Eon was poor in $CO_2$ is as follows: Carbon dioxide, like many other small molecules, forms hydrates or clathrates with the ice from water. 46 molecules of water form a cavity which can contain up to 8 molecules of $CO_2$. This means that 425g of $CO_2$ can be caught up in every kg of $H_2O$ ice. On a frozen earth the pressure conditions for the existence of clathrates can be found at a depth of 100 m in the oceans (Hoffman, N., 2000). The amount of $CO_2$ on the Earth at present is of $3.168 \times 10^{20}$ kg (the majority are carbonate rocks) (Pollak, 1981). If this amount were free, only with 54% of the frozen sea water, the $CO_2$ could be caught in clathrates. But this is the quantity of degassed $CO_2$ in 4.6 thousand millions of years, hence during the Archean there should have been a smaller quantity and a much smaller amount of ice would be necessary to withdraw all the carbon dioxide from the atmosphere. This means that a frozen world would not even have free $CO_2$ in the atmosphere or ocean.

If all the $CO_2$ were to be removed from the atmosphere and dissolved in the ocean the increase in the pH would cause the dissolving of the carbonates rather than its precipitation (Hoffman y Schrag, 2000), but from the previous discussion the conclusion reached is that most of the $CO_2$ was in the clathrates and not dissolved in the sea. Even in the case that some $CO_2$ reached the atmosphere, if some volcanoes unloaded their volatiles in the air, by diffusion in the ice this carbon dioxide would be removed from the atmosphere and form clathrates. Evidence which supports these arguments is a study by Rosing et al. (2010) which shows that the mineralogy of Archean sediments is inconsistent with the presence of large concentrations of $CO_2$ in the Archean atmosphere.

## 5 The ice-albedo feedback

We can overcome the problem of ice-albedo feedback as follows, showing that this is a function of the model and not an intrinsic property of planetary conditions.

5.1 *Temperature model.* In a steady state the energy balance on the planet's surface is:

$$\pi r^2 c(1-A) + 4\pi r^2 e_a \sigma T^4 = 4\pi r^2 e_s \sigma T^4 \qquad (1)$$

The first term is the energy received by the surface of the planet from the Sun. A is the planet's albedo, c is the solar constant, and r is the planet's radius. The second term is the energy, received by the surface of the planet, radiated by the atmospheric layer closest to it. The temperature of this layer is T and its emissivity is $e_a$. Finally the third term is the energy emitted by the planet's surface like a body with a global average emissivity $e_s$, $\sigma$ is the Stefan-Boltzmann constant and T the global average temperature of the earths surface. The model assumes that the temperature of ground and the atmosphere adjacent layer are the same.

If we solve equation (1) for T we have:

$$T = \left[ \frac{c(1-A)}{4(e_s - e_a)\sigma} \right]^{1/4} \quad (2)$$

Taking T=288 K and A= 0.3, the current values for the Earth, and σ= 5.67x10⁻⁸ W m⁻²K⁻⁴ (Curry y Webster, 1999, pp. 437) it is obtained $e_s - e_a$ = 0.61.

In the course of geologic time c is not constant. We have (Kasting and Grinspoon, 1991; Gough, 1981):

$$c = \frac{c_a}{1 + \frac{4}{10}\left(1 - \frac{t}{t_a}\right)} \quad (3)$$

Where $c_a$ is the current solar constant (1370 W/m², Crowley and North, 1991), $t_a$ is the present time measured from the Earth's accretion ($t_a$ = 4.6x10⁹ years = 1.45x10¹⁷ s). The main greenhouse gas on the Earth at present is water vapor. If the planet was frozen almost no vapor was in the atmosphere, which is currently the case with Antarctica, and as a result there was hardly any clouds. So, primitive Earth, had a transparent atmosphere at all frequencies and so $e_a$=0 and the planet albedo was the same as that of the surface. For any material the value of emissivity, i.e. the rate of the radiant energy emitted by a surface to the one emitted by a black body at the same temperature, is equal in value to the absorbtivity, the rate of the radiant energy absorbed by a surface to the total energy received (Eisberg, 1973, pp 56). This is why the absorbtivity value in a material has a direct relation to its albedo and it is equal to 1-A, hence in the case of frozen Earth, $e_s$=1-A.

Substituting these values into the equation (2) the result for frozen Earth is:

$$T = \left[ \frac{c}{4\sigma} \right]^{1/4} \quad (4)$$

*5.2 Model Results*

Substituting the equation (3) into equations (2) and (4) it is possible to obtain the evolution of the temperature over time, using the equation (2) to express the temperatures above freezing point and the equation (4) for temperatures beneath it. The results obtained are shown in figure 1. It can be seen that with equation (4) the defrosting point is reached in a time previous to the present (3620 Ma). In this way it can be seen that the ice-albedo feedback problem is a function of the model used and not an intrinsic phenomenon existing for the planets conditions. The problem of the ice-albedo is probably generated by the lineal approximation which is used by the models which predict it and that are not used here to obtain equation (2). In this model the defrosting point is reached before the end of the Precambrian period (figure 1).

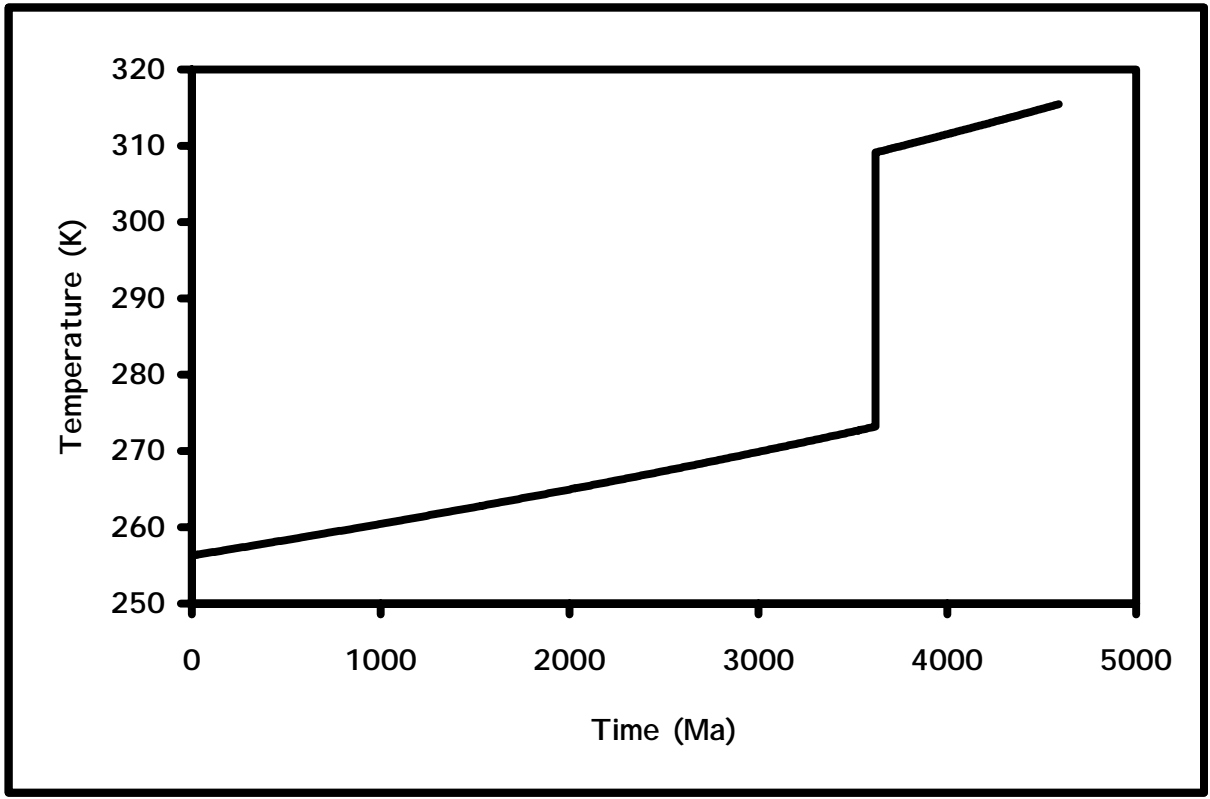

**Figure 1:** Evolution of the Earth temperature with time. The defrosting point was reached at 3620 Ma. In the time before defrosting point it is used equation (4) and after that equation (2). The step in 3620 Ma is due to the change of albedo from a frozen planet to a similar of the current one.

## 6 Evidence of Freezing during the Archean Eon

Evidence of a frozen Earth during the Archean Eon consists of a few glaciogenic sediments (Walker, 1982; Domagal-Goldman et al., 2008). Their scarce abundance can be explained, within the hypothesis of the frozen world, remembering that during this period there was little continental crust and most probably the little there was, was submerged like the actual continental shelf. Actually there was a global ocean with no or very little dry land. If this ocean was frozen, with liquid water under the ice, then no glaciogenic sediments could have been produced due to the fact that submarine rocks were protected from the ice action by this layer of water. And even if the ice reached these rocks the fact of being a global layer prevented any movement and therefore there was no erosion. On the other hand even though some small continents existed, elevated above sea level, the atmosphere would be very dry, due to the fact that all the water was frozen, and no snow would fall and no glacier could form on dry land. On the other hand if the ice extended throughout the ocean, the few glaciers present on the continents of small extension (in fact islands) would have been blocked in their advance by the floating ice hence they would not have eroded the ground.

We know that during the Proterozoic Eon there were glaciations at the beginning and at the end of this Eon. The duration of these glaciations was long, between 2.5 and 2.0 Ga for the first and between 1.0 and 0.57 Ga for the second. The paleo-magnetic data suggests that the final glaciations of the Proterozoic also occurred in equatorial latitudes (Goodwin, 1991). These glaciations were global in all the continents and in all latitudes making the Earth a snowball (Hoffman et al., 1998; Hoffman and Schrag, 2000). This event is evidence of the frozen scenario because during the late Proterozoic the Sun's luminosity is greater than during the Archean. And if in the Archean there had not been any "snowball" hence in the Proterozoic the "snowball" should not have been formed either.

Some evidence shows that the glaciations of the "snowball" Earth during the Proterozoic finished with an extremely high $CO_2$ pressure (Hoffman y Schrag, 2000). In the context of previous discussion this is the expected result and is yet more evidence of the frozen Earth because when the clathrates are melted the $CO_2$ is emitted into the atmosphere. It is highly improbable that these conditions are the product of an accumulative process because the time that these glaciations lasted (~0.5 Ga) is very long and we could expect that in a shorter time the volcanoes could have emitted enough $CO_2$ to produce a greenhouse effect that would have fused the global sea of ice.

Another argument which supports the idea of a frozen Earth is the fact that during Archean times falling of asteroids and comets that produce craters of 20 km or greater were more abundant than in latter periods. These crashes could produce "nuclear winters" similar to the ones produced later with the impact at the end of the Mesozoic. These "nuclear winters" could have triggered the frozen Earth decreasing the planet's temperature below the freezing point. At the beginning of the Archean the Earth was still in the so-called "Heavy Bombardment", the final stage of the planet accretion, and the average time between impacts was around 5,000 years. At the end of the Archean the average period between impacts increased to $1.7 \times 10^6$ years, but it was still lower than at present. The average period of time for impacts which produce craters of 20 km or greater is now $10^7$ years.

We can conclude from the previous arguments, that all presented evidence can support a frozen scenario for the Earth during the Archean. Glaciations in the early Proterozoic and in the late Proterozoic ("snowball" Earth) are only the final stages of this frozen Earth and show that the conditions for a frozen Earth were favorable, at that time.